\documentclass [12pt]{article}
\makeatletter
\def\section{\@startsection {section}{1}{\z@}{-3.5ex plus -1ex minus
 -.2ex}{2.3ex plus .2ex}{\large\bf}}
\def\subsection{\@startsection{subsection}{2}{\z@}{-3.25ex plus -1ex
minus -.2ex}{1.5ex plus .2ex}{\normalsize\bf}}
\makeatother
\makeatletter

\@addtoreset{equation}{section}

\makeatother

\newcommand{\be}{\begin{equation}}
\newcommand{\ee}{\end{equation}}
\newcommand{\bea}{\begin{eqnarray}}
\newcommand{\eea}{\end{eqnarray}}
\def\one{{\rm 1\kern -.9mm l}}

\begin{document}
\begin{titlepage}
\rightline{DFTT 14/2002} \rightline{NORDITA-2002-29 HE}
\vskip 1.8cm
\centerline{\Large \bf ${\cal N}=1$ and ${\cal N}=2$ Super
Yang-Mills theories} \vskip 0.2cm \centerline{\Large \bf from
wrapped branes \footnote{Work partially supported by the European
Commission RTN programme HPRN-CT-2000-00131 and by MIUR under
contract 2001-025492.}} \vskip 1.4cm \centerline{\bf P. Di Vecchia
$^a$, A. Lerda $^{b,c}$ and P. Merlatti $^{c}$} \vskip .8cm
\centerline{\sl $^a$ NORDITA, Blegdamsvej 17, DK-2100 Copenhagen
\O, Denmark} \centerline{\tt divecchia@nbivms.nbi.dk} \vskip .4cm
\centerline{\sl $^b$ Dipartimento di Scienze e Tecnologie
Avanzate} \centerline{\sl Universit\`a del Piemonte Orientale,
I-15100 Alessandria, Italy} \vskip .4cm \centerline{\sl $^c$
Dipartimento di Fisica Teorica, Universit\`a di Torino}
\centerline{\sl and I.N.F.N., Sezione di Torino,  Via P. Giuria 1,
I-10125 Torino, Italy} \centerline{\tt
lerda@to.infn.it~,~merlatti@to.infn.it} \vskip 1.8cm
\begin{abstract}
We consider supergravity solutions of D5 branes wrapped on
supersymmetric 2-cycles and use them to discuss relevant features
of four-dimensional ${\cal N}=1$ and ${\cal N}=2$ super Yang-Mills
theories with gauge group $SU(N)$. In particular in the ${\cal
N}=1$ case, using a gravitational dual of the gaugino
condensate, we obtain the complete NSVZ $\beta$-function. We also
find non-perturbative corrections associated to fractional
instantons with charge $2/N$. These non-perturbative effects
modify the running of the coupling constant which remains finite
even at small scales in a way that resembles to the soft
confinement scenario of QCD.
\end{abstract}
\end{titlepage}
\renewcommand{\thefootnote}{\arabic{footnote}}
\setcounter{footnote}{0} \setcounter{page}{1}
\tableofcontents
\vskip 1cm

\section{Introduction}
Recently, it has become more and more evident that a lot of
relevant information about supersymmetric Yang-Mills (SYM)
theories can be obtained by studying their dual supergravity
backgrounds produced by stacks of D branes. This gauge/gravity
correspondence has been thoroughly investigated in the maximally
supersymmetric and conformal case, where a precise duality can be
established between the ${\cal N}=4$ SYM theory in four dimensions
with gauge group $SU(N)$ and the type IIB supergravity in
$AdS_5\times S_5$ which is the near horizon geometry of $N$ D3
branes in flat space \cite{mal}. Prompted by the remarkable
success of this AdS/CFT duality, a lot of activity has been
devoted to extend the gauge/gravity correspondence also to less
supersymmetric and/or non-conformal theories~\footnote{See the
introduction of Ref.~\cite{d3d7} for a more detailed discussion.}.

One way to reduce the amount of supersymmetry is to place a stack
of D3 branes at the apex of an orbifold \cite{orbi} or of a
conifold \cite{cone}. Depending on the details of the background,
the number of preserved supercharges can be reduced to eight or
four, and thus the SYM theories that correspond to these
configurations will possess ${\cal N}=2$ or ${\cal N}=1$
supersymmetry in four dimensions. Furthermore, both for the
orbifold and the conifold there is a very natural way to break
conformal invariance, namely by means of fractional D3 branes
\cite{frac}. Therefore, in this way one can realize interesting
non-conformal SYM theories in four dimensions with ${\cal N}=2$ or
${\cal N}=1$ supersymmetry (for recent reviews on this approach
see, for example, Refs.~\cite{REV,HKO}).

Another possibility to reduce the number of preserved supercharges
is to consider D branes whose world-volume is partially wrapped on
a supersymmetric cycle inside a K3 manifold or a Calabi-Yau space.
The unwrapped part of the brane world-volume remains flat and
supports a gauge theory. In order to preserve at least some
supersymmetry, the normal bundle to the wrapped D branes has to be
partially twisted \cite{sodov}, and as a consequence of this
twist, some world-volume fields become massive and decouple. This
procedure has been first used in Ref.~\cite{MN} to study the pure
${\cal N}=1$ SYM theory in four dimensions, and later it has been
generalized to many other cases with different space-time
dimensions and different amounts of supersymmetry
\cite{dario,zaffa,wrapped,IMERO,MILANO}~\footnote{We thank W. Lerche for 
pointing out that a different approach to wrapped branes has been considered
in Ref.~\cite{lerche}.}.

In this paper we will use the gauge/gravity correspondence to
study SYM theories with ${\cal N}=1$ and ${\cal N}=2$
supersymmetry in four dimensions. In particular we reconsider in
some detail the supergravity solutions corresponding to $N$
D5-branes wrapped on a 2-sphere that have been first found in
Refs.~\cite{MN} and \cite{dario,zaffa} for the ${\cal N}=1$ and
${\cal N}=2$ case respectively. Even if these two solutions have
been discussed already in several papers, here we review again
their derivation in order to set up the notation, to show their
similarities and differences, and also to be self-contained. In
doing this, we also provide a derivation of the
Maldacena-Nu\~nez (MN) solution of Ref.~\cite{MN} using the first
order formalism~\footnote{A similar derivation in the first
order formalism has been previously presented in Ref.~\cite{tseytlin}.}. 
We then consider the (bosonic) massless
open-string modes that propagate on the flat part of the D5
world-volume and study their effective action at energies where
the higher string modes, as well as the Kaluza-Klein excitations
around the 2-cycles, decouple. This resulting theory is simply a
four-dimensional SYM theory with gauge group $SU(N)$. In both the
${\cal N}=2$ and ${\cal N}=1$ case, it turns out that the coupling
constant $g_{\rm YM}$ and the vacuum angle $\theta_{\rm YM}$ of
this SYM theory can be expressed in terms of the ten-dimensional
supergravity solution representing $N$ wrapped D5 branes in a very
explicit and simple form~\footnote{This situation is very similar
to the one encountered with fractional D3 branes in orbifold
models (see for example
Refs.~\cite{d3,polch,marco,d3d7,FERRO,ANO}).}. In particular, if
we denote by $\rho$ the radial coordinate on which the classical
supergravity fields depend, we find that
\begin{equation}
\frac{1}{g_{\rm YM}^2} = F(\rho) \label{fin1}
\end{equation}
where the function $F(\rho)$ is an explicitly computable
function (see eqs.(\ref{gym2}) and (\ref{gymfin}) below).

The other crucial ingredient for the gauge/gravity correspondence
is the relation between the radial parameter of the supergravity
solution and the energy scales of the gauge theory. For
non-conformal theories in general it may be ambiguous and
difficult to establish such a relation \cite{Peet}; however, for
the two cases under consideration we manage to find a supergravity
realization
of a protected operator of the
gauge theory, so that a definite radius/energy relation can be
obtained~\footnote{Again, this situation is similar to the one of
fractional D3 branes in orbifold models (see, in particular,
Ref.~\cite{ANO}).}. The protected operators we will consider are
the complex scalar field $\Psi$ of the vector multiplet in the
${\cal N}=2$ theory, and the gaugino condensate
$\langle\lambda^2\rangle$ in the ${\cal N}=1$ theory. 
In general, the energy/radius relation that we obtain
in this way takes the form
\begin{equation}
\frac{\mu}{\Lambda} = G(\rho)\label{fin2}
\end{equation}
where $\mu$ is the subtraction point at which the theory is defined,
$\Lambda$ is the scale dynamically generated by quantum corrections
and $G(\rho)$ is an explicit function which depends on the
model (see eqs. (\ref{holo2}) and (\ref{holo1}) below).

Once the functions $F(\rho)$ and $G(\rho)$ are determined, one can
exploit them to find explicit results for the gauge theory. For
example, by eliminating $\rho$ between (\ref{fin1}) and
(\ref{fin2}), one can obtain the running coupling constant and
hence the $\beta$-function. In the ${\cal N}=2$ case, these
manipulations can be performed in an analytic way and
lead to the exact perturbative $\beta$-function. In the ${\cal
N}=1$ case, instead, these manipulations can only be performed
using asymptotic expansions but nevertheless lead to interesting
results. In particular, we will show that the complete NSVZ
$\beta$-function \cite{NSVZ} of the ${\cal N}=1$ theory can be
obtained from the MN solution, once the appropriate energy/radius
relation is enforced. It is remarkable to see that the entire
perturbative $\beta$-function, and not only the 1-loop
approximation, is encoded in a classical supergravity solution!
Actually, the MN solution also shows the presence of
non-perturbative effects in the form of fractional instantons with
charge $2/N$ which smooth out the running of the gauge coupling
constant that remains finite even at small energy. This smooth
behavior as well as the absence of the Landau pole indicate that
the theory softly flows to the confining phase, in analogy to the
soft-confinement scenario of QCD~\cite{soft}. Furthermore, the MN solution
also accounts for the chiral anomaly, the chiral symmetry breaking
and the correct action for the gauge instantons which we manage to
realize explicitly as wrapped D strings.

The analysis of the IR regime of a gauge theory and the study of
its non-perturbative features by means of a supergravity
solution are possible only if the latter is non singular. This is
precisely the case of the MN solution which is regular even at
small distances due the presence of a field that, in
Ref.~\cite{MILANO}, has been identified with the gravitational
dual of the gaugino condensate. In this respect, the MN solution
is very different from the ${\cal N}=2$ solution of
Ref.~\cite{dario,zaffa} which exhibits a naked singularity of
repulson type and an enhan\c{c}on locus \cite{enhancon} that
prevent from obtaining information on the IR regime of the dual
${\cal N}=2$ SYM theory and in particular on its instanton
corrections~\footnote{The same problems occur in the fractional D3
brane solutions~\cite{d3,polch,marco,d3d7}.}. On the contrary, the
MN solution is similar to the Klebanov-Strassler solution
(KS)~\cite{KS} which describes a system of regular and fractional
D3 branes on a deformed conifold. The KS solution is free of
singularities and can be successfully used to describe many
interesting features of a four dimensional ${\cal N}=1$ gauge
theory with chiral matter that, through a series of duality
cascades, eventually flows to the pure ${\cal N}=1$ SYM theory in
the deep IR to which our analysis can be extended~\cite{Emil}. A
qualitative study of the general implications of the MN and KS
solutions for ${\cal N}=1$ gauge theories can be found in
Ref.~\cite{LS}.

The paper is organized as follows. In Section 2 we review the
${\cal N}=2$ supergravity solution of Ref.~\cite{dario,zaffa} and
the ${\cal N}=1$ solution Ref.~\cite{MN} in a first order
formalism. In Section 3 we discuss the pure ${\cal N}=2$ SYM
theory using the dual supergravity solution of Section 2.1. In
particular, we derive the running of the coupling constant and the
chiral anomaly, but differently from Ref.~\cite{dario} we do not
use the probe technique. In Section 4 we study the pure ${\cal
N}=1$ SYM theory and, by exploiting the supergravity solution
of Section 2.2, we discuss the chiral anomaly, the gaugino
condensate, the running coupling constant, the NSVZ
$\beta$-function and the instanton action. Finally, in Section 5
we present our conclusions, while in the Appendix we give some
technical details about the parameterizations used to derive the
supergravity solutions of Section 2.
\vskip 1.5cm
\section{The Supergravity Solutions}

In this section we review the type IIB supergravity solutions that
correspond to ${\cal N}=1$ and ${\cal N}=2$ SYM theories in four
dimensions with gauge group $SU(N)$ and no matter.

The starting point is a stack of $N$ D5 branes with two
longitudinal directions wrapped on a supersymmetric 2-cycle. The
difference between ${\cal N}=1$ and ${\cal N}=2$ arises from the
way in which the 2-cycle is embedded in the ambient space, a
Calabi-Yau threefold for ${\cal N}=1$ and a K3 manifold for ${\cal
N}=2$. The four unwrapped longitudinal directions of the D5 branes
span a flat world-volume where the SYM theory lives. In order to
preserve the proper amount of supersymmetry the normal bundle to
the five-branes has to be topologically twisted \cite{sodov}; this is
achieved by embedding the $U(1)$ spin-connection of the 2-cycle
into $SO(4)$ which is the R-symmetry group of the D5 branes. As
remarked in \cite{MNuno,MN}, an efficient way to derive the
geometry corresponding to this set-up is to consider the $SO(4)$
gauged supergravity in seven dimensions, find domain-walls wrapped
on a 2-sphere, and then lift them up in ten dimensions.

The fields of the $SO(4)$ gauged supergravity in $d=7$ are the
metric $G_{\mu\nu}$, the gauge fields $A_\mu^{ij}=-A_\mu^{ji}$
with $i,j=1,\ldots,4$ being vector indices of $SO(4)$, a symmetric
matrix $T_{ij}$ of scalars and a 2-form potential
$A^{(2)}_{\mu\nu}$ \cite{sezgin}. The Lagrangian of this
supergravity theory is
\begin{eqnarray}
L_{7}&=& \sqrt{-\det G} \,\left\{ R(G) - \frac{5}{16}\,\partial_\mu
y \,\partial^\mu y -\frac{1}{4}\, {\widetilde T}_{ij}^{-1}{\cal
D}_\mu{\widetilde T}_{jk}\, {\widetilde T}_{k\ell}^{-1}{\cal
D}^\mu{\widetilde T}_{\ell i} \right. \nonumber
\\
&&\left. -\,\frac{1}{8}\,{\rm e}^{-y/2}\,{\widetilde
T}_{ik}^{-1}{\widetilde
T}_{j\ell}^{-1}{F}^{ij}_{\mu\nu}{F}^{k\ell\,\mu\nu}
-\frac{1}{12}\,{\rm e}^{-y} \,H_{\mu\nu\rho}H^{\mu\nu\rho} -V
\right\} \label{L7}
\end{eqnarray}
where
\begin{equation}
{\rm e}^{y}= \det T ~~~,~~~{\widetilde T}_{ij}= {\rm e}^{-y/4}\,{
T}_{ij}~~, \label{ttilde}
\end{equation}
and
\begin{eqnarray}
{\cal D}_\mu{\widetilde T}_{ij}&=&
\partial_\mu {\widetilde T}_{ij}
+\lambda\left(A_\mu^{ik}{\widetilde T}_{kj}+A_\mu^{jk}{\widetilde
T}_{ik} \right)~~,\label{covder}
\\{F}^{ij}_{\mu\nu} &=& \partial_\mu A_\nu^{ij} -
\partial_\nu A_\mu^{ij}
+\lambda\left(A_\mu^{ik}A_\nu^{kj}-A_\nu^{ik}A_\mu^{kj}\right)~~,
\label{fmunu}\\ V&=& \frac{\lambda^2}{2}\,{\rm e}^{y/2}
\left(2{\widetilde T}_{ij}{\widetilde T}^{ij}-({\widetilde
T}_{ii})^2\right)~~.\label{pot}
\end{eqnarray}
Finally, $H_{\mu\nu\rho}$ are the components of the following
3-form
\begin{equation}
H^{(3)}=dA^{(2)}+\frac{1}{8}\epsilon_{ijk\ell}\left( F^{ij}\wedge
A^{k\ell} -\frac{\lambda}{3} A^{ij}\wedge A^{km}\wedge A^{m\ell}
\right)~~. \label{H}
\end{equation}
In these equations $\lambda$ denotes the $SO(4)$ gauge coupling
constant which has dimensions of a (length)$^{-1}$.

As mentioned above, we are interested in domain-walls of this
supergravity theory that are wrapped on a 2-sphere. Thus we look
for metrics of the form
\begin{equation}
d s_{7}^{2} ={\rm e}^{2f(r)} \left( d x_{1,3}^2 + dr^2 \right) +
\frac{1}{\lambda^2}\,{\rm e}^{2 g(r)}\,d\Omega_2^2 \label{anme}
\end{equation}
where $d x_{1,3}^2$ is the Minkowski metric on ${\rm I\!R}_{1,3}$,
$r$ is the transverse coordinate to the domain-wall, and
$d\Omega_2^2 = d \widetilde\theta^2 + \sin^2 \widetilde\theta\,
d\widetilde\varphi^2$ (with $0\leq \widetilde\theta\leq\pi$ and
$0\leq\widetilde\varphi \leq2\pi$) is the metric of a unit
2-sphere~\footnote{Notice that the factor of $\lambda^{-2}$ in
(\ref{anme}) is necessary for dimensional reasons.}. To fully
determine the domain-wall configuration we must  specify also the
profile of the other supergravity fields. However, to do this we
have to distinguish between the ${\cal N}=1$ and ${\cal N}=2$
cases which require different types of Ansatz. In the next
subsection we review the ${\cal N}=2$ case \cite{dario,zaffa},
while the ${\cal N}=1$ case \cite{MN} will be considered in the
last subsection.

\subsection{The ${\cal N}=2$ Solution}

In order to obtain a solution corresponding to a gauge theory with eight
supercharges in four dimensions, one must first truncate the
$SO(4)$ gauge group of the $d=7$ supergravity to its diagonal
$U(1)$ subgroup, {\it i.e.} to the abelian part of the diagonal
$SU(2)_D\subset SO(4)$. This truncation can be achieved, for
example, by setting
\begin{equation}
A_\mu^{12}\equiv -\,A_\mu \label{a}
\end{equation}
with all other $A^{ij}_\mu$'s put to zero. The consistency of this
abelian reduction requires that the unimodular matrix ${\widetilde
T}_{ij}$ takes the form~\cite{CVETIC}
\begin{equation}
{\widetilde T}_{ij} = {\rm diag}(e^x , e^x , e^{-x} , e^{-x} )~~.
\label{tij5}
\end{equation}
With these positions it is easy to realize that the Lagrangian
(\ref{L7}) becomes
\begin{eqnarray}
L_7 &=& \sqrt{-\det G}\, \left\{ R(G) -
\frac{5}{16} \,\partial_{\mu} y \,\partial^{\mu} y
- \partial_{\mu} x \,\partial^{\mu} x - \frac{1}{4}\,{\rm e}^{-2x - y/2}
F_{\mu \nu}F^{\mu \nu} \right.\nonumber\\
&&\left.-\,\frac{1}{12}\,{\rm e}^{-y} \,H_{\mu\nu\rho}H^{\mu\nu\rho}
+ 4 \,\lambda^2 \,{\rm e}^{y/2}\,  \right\}
\label{l7g}
\end{eqnarray}
where $F_{\mu \nu}=\partial_{[\mu}
A_{\nu]}$ and $H_{\mu\nu\rho}=\partial_{[\mu}
A^{(2)}_{\nu\rho]}$.

We now look for solutions of the field equations following from
the Lagrangian (\ref{l7g}) in which the metric has the form
(\ref{anme}) and the fields $x$ and $y$ are functions only of the
transverse coordinate $r$. To implement the topological twist that
preserves eight supercharges, we must identify the $U(1)$ gauge
field with the spin-connection on the 2-sphere and thus write
\begin{equation}
A = -\, \frac{1}{\lambda}\cos \widetilde\theta\,d\widetilde\varphi
\label{af56}
\end{equation}
where the coupling constant $\lambda$ has been introduced for
dimensional reasons. Furthermore, we can consistently set
$A^{(2)}=0$.

The task is then reduced to find the four functions $f(r)$,
$g(r)$, $x(r)$ and $y(r)$ that solve the field equations. This has
been done explicitly in Ref.~\cite{dario} \footnote{See also
Ref.~\cite{zaffa}.} where it has been shown that it is consistent
to impose the relation $y= -4 f$ and that, after defining
\begin{equation}
h = g - f~~~,~~~ k = \frac{3}{2}\, f + g~~, \label{rel98}
\end{equation}
it is possible to derive the field equations for $h(r)$, $k(r)$
and $x(r)$ from the auxiliary Lagrangian
\begin{equation}
{\cal L} = {\rm e}^{2k} \left[ 4 \dot{k}^2 - 2 \dot{h}^2 -
\dot{x}^2 - {\cal V} \right] \label{lagra33}
\end{equation}
where dots denote derivative with respect to $r$ and
\begin{equation}
{\cal V} =  - 4 \lambda^2 -2\lambda^2 {\rm e}^{-2h}  +
\frac{\lambda^2}{2} {\rm e}^{-4h -2 x} \label{pote49}
\end{equation}
together with the additional condition that the Hamiltonian ${\cal H}$
associated to ${\cal L}$ vanishes. This condition, which
reads
\begin{equation}
{\cal H} = \frac{1}{4}\,{\rm e}^{-2k}
\left(\frac{1}{4}\,p_k^2-\frac{1}{2}\,p_h^2 -p_x^2\right)
+{\rm e}^{2k}\,{\cal V} = 0~~,
\label{hamil}
\end{equation}
is a signal of the fact that the second-order field equations can
be derived from a system of first-order (or BPS) equations, as one
can show with a Hamilton-Jacobi approach \cite{ferretti}. In fact,
by introducing the principal function ${\cal F}(k,h,x)$ such that
$ p_k = \frac{\partial {\cal F}}{\partial k}$, $p_h=
\frac{\partial {\cal F}}{\partial h}$, $p_{x} = \frac{\partial
{\cal F}}{\partial x}$ and assuming that ${\cal F} (k,h,x) = {\rm
e}^{2k} {\cal W}(h,x)$, it is easy to see that (\ref{hamil})
becomes
\begin{equation}
\frac{1}{8} \left(\frac{\partial {\cal W}}{\partial h} \right)^2 +
\frac{1}{4} \left(\frac{\partial {\cal W}}{\partial x} \right)^2
 - \frac{1}{4}\, {\cal W}^2 = {\cal V}~~,
\label{eqw99}
\end{equation}
which is solved by
\begin{equation}
{\cal W} = -4 \lambda \cosh x - \lambda {\rm e}^{ - 2h-x}~~.
\label{solw84}
\end{equation}
The remaining Hamilton equations yield the following first-order
system
\begin{equation}
\dot{k} = \frac{1}{4}\, {\cal W} ~~,~~
\dot{h} = -\frac{1}{4}\,\frac{\partial{\cal W}}{\partial h}~~
,~~\dot{x} =-\frac{1}{2} \, \frac{\partial {\cal W}}{\partial x}~~,
\label{first64}
\end{equation}
which, as shown in Ref.~\cite{dario}, can be conveniently solved
in terms of the variable $z \equiv {\rm e}^{2h}$. In fact, one finds
\begin{equation}
{\rm e}^{-2x} = 1 - \frac{1+ c\, {\rm e}^{-2 z}}{2 z}~~~,~~~
{\rm e}^{2k +x} = z\, {\rm e}^{2 z}
\label{sol63}
\end{equation}
where $c$ is an arbitrary integration constant. Thus, choosing $z$ as new
radial coordinate, the functions appearing in the domain-wall solution
are
\begin{equation}
f(z)=\frac{2}{5}\,z - \frac{1}{5}\,x(z)~~,~~g(z)=f(z)+\frac{1}{2}\,\log(z)
~~,~~y(z)= -4f(z)
\label{solz}
\end{equation}
with $x(z)$ given in (\ref{sol63}).

We now up-lift this solution to ten dimensions in order to exhibit
it as a D5 brane configuration of Type IIB supergravity. This
up-lift can be explicitly performed using the formulas of
Ref.\cite{CVETIC} which lead to a metric, a dilaton and a 2-form
of magnetic type in ten dimensions. In this respect, however, we
would like to point out that in Ref.~\cite{CVETIC} only the
up-lift to NS5 brane configurations is considered. Indeed, the
magnetic 2-form that is produced in ten dimensions has to be
identified with the NS-NS antisymmetric tensor $B_{\mu\nu}$ in
view of the sign that is chosen for the exponential coupling with
the dilaton in the Einstein frame. However, it is straightforward
to adjust the up-lift formulas of Ref.~\cite{CVETIC} to the other
sign and thus produce a magnetic 2-form which can be identified
with a R-R potential $C^{(2)}$ in ten dimensions. In this way, the
seven-dimensional solution can be directly up-lifted to a D5 brane
configuration of type IIB, which is what one needs to discuss the
correspondence with a dual SYM theory in four dimensions.

With this in mind, we parameterize the 3-sphere that leads from
seven to ten dimensions with the angles $\theta$, $\phi_1$ and
$\phi_2$ (with  $ 0\leq \theta  \leq \frac{\pi}{2} $ and $ 0 \leq
\phi_{1,2} \leq 2 \pi$) as discussed in Appendix A (see
eq.(\ref{sphere3})), and use the (modified) up-lift formulas of
Ref.~\cite{CVETIC} to find a ten-dimensional (string frame) metric
given by
\begin{eqnarray}
ds_{10}^2 &=& {\rm e}^{\Phi}\Bigg[ d x_{1,3}^2 +
\frac{z}{\lambda^2} \left( d {\widetilde{\theta}}^2 +
\sin^2 {\widetilde{\theta}} \,d {\widetilde{\varphi}}^2 \right)
+\frac{1}{\lambda^2} \,{\rm e}^{2x} \,dz^2 \label{met10} \\
&&~
+ \frac{1}{\lambda^2} \left(d \theta^2 +
\frac{{\rm e}^{-x}}{\Omega} \cos^2 \theta \left(d
  \phi_{1} + \cos {\widetilde{\theta}} \,d {\widetilde{\varphi}}
\right)^2
+ \frac{{\rm e}^{x}}{\Omega} \sin^2 \theta \,d \phi_{2}^{2}
 \right)\Bigg]~~,
\nonumber
\end{eqnarray}
a dilaton given by
\begin{equation}
{\rm e}^{2\Phi} = {\rm e}^{2 z} \left[1 - \sin^2 \theta
~\frac{1+c \,{\rm e}^{-2 z}}{2  z} \right]~~,
\label{dil}
\end{equation}
and a magnetic R-R 2-form given by
\begin{equation}
C^{(2)} = \frac{1}{\lambda^2}\, \phi_2\,d \left[\frac{\sin^2
\theta}{\Omega\,{\rm e}^{x}} \,( d \phi_1 + \cos
{\widetilde{\theta}}\, d {\widetilde{\varphi}}) \right]
\label{b2}
\end{equation}
where
\begin{equation}
\Omega = {\rm e}^{x} \cos^2 \theta + {\rm e}^{-x} \sin^2 \theta~~.
\label{omega45}
\end{equation}
We remark that this D5 brane solution agrees with the one of
Ref.~\cite{dario} which is obtained by performing a S-duality
transformation on a NS5 brane configuration. However, the metric
(\ref{met10}) is written in a way in which the role of the
different coordinates and factors is not immediately clear. Thus,
in analogy with what has been done in Ref.~\cite{IMERO}, we
perform the following change of variables
\begin{equation}
\rho = {\sin \theta } \,{\rm e}^{ z}~~,~~
\sigma = {\sqrt{z}} \,\cos \theta \,{\rm e}^{ z- x}
\label{new49}
\end{equation}
and rewrite eq. (\ref{met10}) as follows
\begin{eqnarray}
ds_{10}^2 &=& H^{-1/2} \left[d x_{1,3}^{2} + \frac{z}{\lambda^2}
\left( d {\widetilde{\theta}}^2 + \sin^2 {\widetilde{\theta}}\,
  d{\widetilde{\varphi}}^2 \right)\right]
+\frac{H^{1/2}}{\lambda^2}\left( d \rho^2 + \rho^2 d \phi_{2}^{2}\right)
\nonumber \\
&&
+~\frac{H^{1/2}}{\lambda^2\,z} \left[ d\sigma^2 + \sigma^2
\left(d \phi_1 + \cos {\widetilde{\theta}} \,d {\widetilde{\varphi}}
 \right)^2 \right]
\label{metfin}
\end{eqnarray}
where $H\equiv {\rm e}^{-2\Phi}$. In this form the structure of
the metric is much clearer. In fact, one can distinguish the
transverse space into a plane parameterized by $\rho$ and
$\phi_2$ (see the last two terms in the first line of
(\ref{metfin})), and a non-trivial twisted plane pertaining to the
K3 manifold (see the second line of (\ref{metfin})). Thus one can say that
the two (dimensionless)
coordinates $\rho$ and $\sigma$ defined in (\ref{new49})
represent two radial directions, respectively in the flat
transverse space and in the curved space transverse to the brane
but non-trivially fibered on the 2-cycle along which the brane is
wrapped. In the next section we will show that the flat radial
coordinate $\rho$ is directly related to the energy scale of the
dual four dimensional gauge theory, and that shifts in the angle
$\phi_2$ are directly related to chiral transformations.

We conclude this subsection by observing that the R-R charge of
the configuration (\ref{met10})-(\ref{b2}) is given by
\begin{equation}
Q_5 \,=\, \frac{1}{2\kappa_{10}^2}\int_{S_3}dC^{(2)} \,=\,
\frac{2\pi^2}{\kappa_{10}^2\,\lambda^2} \label{charge}
\end{equation}
where $\kappa_{10}=8\pi^{7/2}g_s\alpha'^2$ is the gravitational
coupling constant of the Type IIB string theory and $S_3$ is the
transverse 3-sphere at infinity.
Since this charge must be an integer multiple
of the elementary D5 charge
\begin{equation}
\tau_5=(2\pi)^{-5}g_s^{-1}\alpha'^{-3}~~,
\label{tau5}
\end{equation}
we deduce that the seven-dimensional coupling constant
$\lambda$ can be written in terms of string theory parameters as
follows
\begin{equation}
\frac{1}{\lambda^2}= N\,g_s\,\alpha' \label{lambda}
\end{equation}
where $N$ is the number of wrapped D5 branes.

\subsection{The ${\cal N}=1$ Solution}

We now discuss the MN supergravity solution of Ref.~\cite{MN}
which corresponds to a gauge theory with four supercharges in four
dimensions. As before, we start from the $SO(4)$ gauged
supergravity Lagrangian (\ref{L7}), but this time we gauge one of
the two $SU(2)$ factors inside $SO(4)$ (say, for example,
$SU(2)_L$). This is achieved by imposing the anti-selfduality
constraint
\begin{equation}
A_\mu^{ij}=-\frac{1}{2} \,\epsilon^{ijk\ell}A_\mu^{k\ell}
\label{a1}
\end{equation}
which explicitly reads
\begin{equation}
A_\mu^{34}=-A_\mu^{12}\equiv
A_\mu^{3}~~,~~A_\mu^{24}=-A_\mu^{31}\equiv
A_\mu^2~~,~~A_\mu^{14}=-A_\mu^{23}\equiv A_\mu^1~~,\label{a2}
\end{equation}
The vectors $A_\mu^a$ (with $a=1,2,3$) are the gauge fields of
$SU(2)_L$ whose field strengths are $F^a=d A^{a} +
\lambda~\varepsilon^{abc} A^{b} \wedge A^{c}$ as one can see by
inserting (\ref{a1}) and (\ref{a2}) into (\ref{fmunu}). The
consistency of this $SU(2)_L$ reduction requires that the
unimodular matrix ${\widetilde T}_{ij}$ be simply~\cite{CVETIC}
\begin{equation}
{\widetilde T}_{ij}=\delta_{ij}~~.
\label{t1}
\end{equation}
With these positions the Lagrangian
(\ref{L7}) becomes
\begin{eqnarray}
L_7 &=& \sqrt{-\det G}\, \left\{ R(G) - \frac{5}{16}
\,\partial_{\mu} y \,\partial^{\mu} y - \frac{1}{2}\,{\rm e}^{-
y/2} F^a_{\mu \nu}F^{a\,\mu \nu} \right.\nonumber\\
&&\left.-\,\frac{1}{12}\,{\rm e}^{-y}
\,H_{\mu\nu\rho}H^{\mu\nu\rho} + 4 \,\lambda^2 \,{\rm e}^{y/2}\,
\right\} \label{l7g2}
\end{eqnarray}
where $H_{\mu\nu\rho}$ are the components of the 3-form
\begin{equation}
H^{(3)}=dA^{(2)}+ F^{a}\wedge
A^{a} -\frac{\lambda}{12}\,\epsilon^{abc}\,A^{a}\wedge A^{b}\wedge A^{c}
~~. \label{H1}
\end{equation}

Like in the ${\cal N}=2$ case, we look again for domain-wall
solutions where the metric is of the form (\ref{anme}) and the
scalar $y$ is function only of the radial coordinate $r$. To
implement the topological twist that preserves four supercharges
we should identify the spin-connection on the 2-sphere with the
gauge field of a $U(1)\subset SU(2)_L$. If we simply do this, the
corresponding supergravity solution turns out to be singular and unphysical.
However, as discovered in Ref.~\cite{MN}, the singularity can be smoothed out
by considering a more general Ansatz in which all gauge fields of
$SU(2)_L$ are switched-on, namely by taking \cite{CV}
\begin{equation}
A^{1} = -\,\frac{1}{2\lambda}\, a(r)\, d{\widetilde{\theta}}~~,~~
A^{2} =  \frac{1}{2\lambda}\,a(r) \sin {\widetilde{\theta}}\, d
{\widetilde{\varphi}}~~,~~ A^{3} = -\,\frac{1}{2\lambda}\,\cos
{\widetilde{\theta}}\, d
 {\widetilde{\varphi}}~~.
\label{vec45}
\end{equation}
With this Ansatz it is easy to realize that the only non-vanishing
components of the field strengths are
\begin{equation}
F^{1}_{r {\tilde{\theta}}} = -\,\frac{1}{2\lambda}\,{\dot{a}}
~~,~~ F^{2}_{r {\tilde{\varphi}}} = \frac{1}{2\lambda}\,{\sin
{\widetilde{\theta}}\,{\dot{a}}}~~,~~ F^{3}_{{\tilde{\theta}}
{\tilde{\varphi}}} = \frac{1}{2\lambda}\,{\sin
{\widetilde{\theta}}}\left(1 - a^2 \right)~~. \label{vec46}
\end{equation}
Moreover one finds that $F^{a}\wedge A^{a}=A^{1}\wedge A^{2}\wedge
A^{3}=0$ so that the 3-form (\ref{H1}) simply becomes
$H^{(3)}=dA^{(2)}$. Therefore, it is consistent to set $A^{(2)}=0$
and look for a classical configuration that depends only on the
functions $f(r)$, $g(r)$, $y(r)$ and $a(r)$.

Inserting this Ansatz into the supergravity field equations that follow
from (\ref{l7g2}), we find that it is possible to set $y = -4 f$ and
derive the remaining equations from the auxiliary Lagrangian
\begin{equation}
{\cal L} = {\rm e}^{2k} \left[ 4 \dot{k}^2 - 2 \dot{h}^2 -
\frac{\dot{a}^2}{2} {\rm e}^{-2h} - {\cal V}  \right]
\label{lagra32}
\end{equation}
where $h$ and $k$ are defined as in (\ref{rel98}) and
\begin{equation}
{\cal V} =  - 4 \lambda^2 -2\lambda^2 \,{\rm e}^{-2h}  +
\frac{\lambda^2\,(1 - a^2)^2}{4} \,{\rm e}^{-4h} \label{pote48}
\end{equation}
provided we require that the Hamiltonian ${\cal H}$
associated to ${\cal L}$ vanishes.
This condition explicitly reads
\begin{equation}
{\cal H} = \frac{1}{2}\,{\rm e}^{-2k}
\left(\frac{1}{8}\,p_k^2-\frac{1}{4}\,p_h^2 -{\rm
e}^{2h}\,p_a^2\right) +{\rm e}^{2k}\,{\cal V} = 0~~.
\label{hamil1}
\end{equation}
We now proceed as in the ${\cal N}=2$ case with the
Hamilton-Jacobi method and introduce the principal function ${\cal
F}(k,h,a)$ such that $p_k = \frac{\partial {\cal F}}{\partial k}$,
$p_h =\frac{\partial {\cal F}}{\partial h}$, $p_{a} =
\frac{\partial {\cal F}}{\partial a}$. Then, if we assume that
${\cal F}(k,h,a) = {\rm e}^{2k}\,{\cal W}(h,a)$, eq.
(\ref{hamil1}) becomes
\begin{equation}
\frac{1}{8} \left(\frac{\partial {\cal W}}{\partial h} \right)^2 +
\frac{1}{2} \left(\frac{\partial {\cal W}}{\partial a} \right)^2
{\rm e}^{2h} - \frac{1}{4} \,{\cal W}^2 = {\cal V}~~,
\label{eqw98}
\end{equation}
which is solved by (see also Ref.~\cite{tseytlin})
\begin{equation}
{\cal W} = \lambda \,{\rm e}^{-2h} \, \sqrt{(1 + 4 {\rm e}^{2h}
)^2 + 2(-1 + 4  {\rm e}^{2h}) a^2 + a^4}~~. \label{solw89}
\end{equation}
Having the explicit form of ${\cal W}$, the problem is reduced to
a system of first-order (BPS) equations given by
\begin{equation}
\dot{k} =\frac{1}{4}\, {\cal W}~~,~~  \dot{h} = -\frac{1}{4}\,
\frac{\partial {\cal W}}{\partial h}~~,~~ \dot{a} = -\, {\rm
e}^{2h}\, \frac{\partial {\cal W}}{\partial a} \label{first67}
\end{equation}
which can be explicitly solved. Indeed one finds
\begin{eqnarray}
{\rm e}^{2h} &=&  \rho \coth 2 \rho - \frac{\rho^2}{\sinh^2 2
\rho} - \frac{1}{4}~~, \label{solh6}\\ {\rm e}^{2k} &=&  {\rm
e}^{h}\,\frac{\sinh 2\rho}{2}~~, \label{solh61}\\
a &=& \frac{2
\rho}{\sinh 2 \rho} \label{solh62}
\end{eqnarray}
where $\rho \equiv \lambda \,r$. This solution has been first
found in Ref.~\cite{CV}, although in a different context. The
functions originally appearing in the domain-wall solution are
then
\begin{equation}
f(\rho)=\frac{1}{5}\,\log\left(\frac{\sinh 2\rho}{2\,{\rm
e}^h}\right) ~~,~~g(\rho)=f(\rho)+h(\rho)~~,~~y(\rho)=-4f(\rho)
\label{sol80}
\end{equation}
with $h(\rho)$ given in (\ref{solh6}). It is interesting to
observe that $a\sim\rho\,{\rm e}^{-2\rho}$ for $\rho\to\infty$;
thus only the $A^3$ component of the $SU(2)_L$ gauge field
effectively survives in the large $\rho$ region and the gauge
group reduces to $U(1)$ .

We now up-lift this solution to ten dimensions to exhibit it as a
D5 brane configuration of Type IIB supergravity. To this aim we
could choose the same parameterization of the 3-sphere that we
used in the previous subsection when we up-lifted the ${\cal N}=2$
solution. However, for the applications to the dual gauge theory
that we will discuss in the following, it is more convenient to
choose a different parameterization and describe the 3-sphere with
the Euler angles $\theta'$, $\phi$ and $\psi$ (with
$0\leq\theta'\leq \pi$, $0\leq\phi\leq 2\pi$ and $0\leq\psi\leq
4\pi$) as described in (\ref{para49}), and use the corresponding
left-invariant 1-forms $\sigma^a$ given in (\ref{diffe49}). Then,
using the (modified) up-lift formulas of Ref.~\cite{CVETIC}, we
find a ten-dimensional (string frame) metric
\begin{equation}
d s_{10}^{2}
={\rm e}^\Phi
\left[  dx_{1,3}^{2} +
  \frac{{\rm e}^{2h}}{\lambda^2} \left(d {\widetilde{\theta}}^2
+ \sin^2 {\widetilde{\theta}}\,d {\widetilde{\varphi}}^2 \right)
\right]
+\frac{{\rm e}^\Phi
}{\lambda^2} \left[ d \rho^2 +
\sum_{a=1}^3\left( \sigma^a - \lambda A^a \right)^2\right]~~,
\label{stri987}
\end{equation}
a dilaton
\begin{equation}
{\rm e}^{2 \Phi} =
\frac{\sinh 2 \rho}{2 \,{\rm e}^h}~~,
\label{dila493}
\end{equation}
and a magnetic R-R 2-form
\begin{eqnarray}
{C}^{(2)} &=& \frac{1}{4 \lambda^2} \left[ \psi\left(\sin
\theta'\, d \theta'
  \wedge d \phi - \sin {\widetilde{\theta}} \,d {\widetilde{\theta}}
  \wedge d {\widetilde{\varphi}}
\right) - \cos \theta' \,\cos
  {\widetilde{\theta}} \,
d \phi \wedge d {\widetilde{\varphi}}\right]\nonumber \\
&&+\,\frac{a}{2\lambda^2} \left[ d{\widetilde{\theta}} \wedge
\sigma^1 - \sin {\widetilde{\theta}}\,
 d{\widetilde{\varphi}} \wedge \sigma^2 \right]
\label{f3exact}
\end{eqnarray}
with field strength
\begin{equation}
F^{(3)} = \frac{2}{\lambda^2} \,
\left(\sigma^1 - \lambda
A^{1}\right) \wedge \left(\sigma^2 - \lambda A^{2}\right) \wedge
\left(\sigma^3 -\lambda A^{3}\right) - \frac{1}{\lambda}\,
\sum_{a=1}^3F^{a} \wedge \sigma^a~~. 
\label{h3f39}
\end{equation}
By computing the R-R charge $Q_5$ of this configuration 
(see eq.(\ref{charge})) and imposing its quantization in 
units of the elementary D5 charge
$\tau_5$, we find again that
\begin{equation}
\frac{1}{\lambda^{2}}=N\,g_s\,\alpha'
\label{lambda1}
\end{equation}
where $N$ is the number of wrapped five-branes. The solution
(\ref{stri987})-(\ref{h3f39}) entirely agrees with the one of
Ref.\cite{MN}.

We conclude this section by observing that there are many common
features and similarities between the ten-dimensional ${\cal N}=2$
and ${\cal N}=1$ solutions as we have presented them here, despite
the fact that we have used different angular variables in the two
cases. After all, this is not too surprising since both solutions
originate from the same Ansatz (\ref{anme}) in seven dimensions;
but nevertheless it is noteworthy to see that in the first-order
formalism the two solutions are very similar and that a flow
between them could be in principle established in terms of the
``superpotentials'' ${\cal W}$ defined in eqs.(\ref{solw84}) and
(\ref{solw89}). Of course there are also many important
differences between the two solutions. In particular we would like
to emphasize that in the ${\cal N}=2$ case one can distinguish a
plane in the transverse space where the wrapped D5 branes can
move, while in the ${\cal N}=1$ case no such plane exists.
Moreover the ${\cal N}=2$ solution is singular at short distance
while the ${\cal N}=1$ solution is regular. These facts have
important consequences for the dual SYM theories as we shall
discuss in the next sections.

\vskip 1.5cm
\section{The ${\cal N}=2$ Super Yang-Mills Theory}

The supergravity solution presented in subsection 2.1 describes
the geometry produced by $N$ D5 branes wrapped on a supersymmetric
2-cycle in such a way that one quarter of the 32 supercharges of Type IIB
are preserved. The four-dimensional part of the D5 world-volume
that is not involved in the wrapping process remains
flat and is the Minkowski space-time ${\rm I\!R}_{1,3}$
where a supersymmetric gauge theory with 8 supercharges is
defined. To specify this theory one must determine which massless open-string
modes can propagate in ${\rm I\!R}_{1,3}$. By
applying for example the methods of Ref.~\cite{sodov}, one can find that
these modes fill a ${\cal N}=2$ vector multiplet in the adjoint
representation of $SU(N)$, and thus the low-energy four-dimensional
gauge theory is a pure ${\cal N}=2$ SYM theory with gauge group $SU(N)$.

We now show how the supergravity solution (\ref{dil}), (\ref{b2})
and (\ref{metfin}) can be used to extract information on the corresponding
gauge theory. First of all, we observe that in the transverse
space to the D5 branes one can select a two-dimensional sub-space
in which the branes can freely move. This is the locus $\sigma=0$
and thus is the plane parameterized by $\rho$ and $\phi_2$. To
find this result we can use the ``probe technique''
\cite{primer,REV} and study the dynamics of a wrapped D5 brane in
the background geometry (\ref{dil})-(\ref{metfin}). The (string
frame) world-volume action for a probe D5 brane is
\begin{equation}
S = - \tau_5 \int d^6 \xi ~{\rm e}^{- \Phi}
\sqrt{- \det \left( G+ 2 \pi \alpha' F \right)} + \tau_5
\int\left(\sum_n C^{(n)}\wedge {\rm e}^{2 \pi \alpha' F}
\right)_{6-{\rm form}}\label{borni78}
\end{equation}
where $\tau_5$ is defined in (\ref{tau5}), $F$ is a world-volume
gauge field and all bulk fields are understood to be the
pull-backs onto the brane world-volume which is parameterized by
$\xi=\{x^0,x^1,x^2,x^3,\widetilde \theta,\widetilde\varphi\}$. By
expanding the action (\ref{borni78}) in powers of $\alpha'$ and
substituting in it the solution (\ref{dil}), (\ref{b2}) and
(\ref{metfin}), we find that there is a static potential between
the probe and the source given by
\begin{equation}
- \tau_5\, \int d^6 \xi\,\frac{z}{\lambda^2 H}\,
\sin {\widetilde{\theta}}
\left[ \left( 1 +
\frac{\lambda^2 H \sigma^2}{ z^2 \tan^2 {\widetilde{\theta}}}
\right)^{1/2} -1 \right]~~.
\label{nofor65}
\end{equation}
In order to satisfy the ``zero-force'' condition and hence describe
a BPS supersymmetric configuration, this potential term should vanish.
Thus, we should require that the probe brane be placed at $\sigma=0$.
This means that in order to preserve supersymmetry the branes cannot move
arbitrarily in their transverse space but only in the locus $\sigma=0$,
{\it i.e.} in the $(\rho,\phi_2)$-plane.

From now on, we abandon the probe approach but still we set
$\sigma=0$. Then, the relevant part of the transverse D5 brane
metric (\ref{metfin}) can be simply written as
\begin{equation}
ds^2_{\rm transv} = H^{1/2} d{\overline Z}\,dZ
\label{mettrans}
\end{equation}
where we have defined
\begin{equation}
Z= \frac{1}{\lambda}\,\rho\,{\rm e}^{{\rm i}\phi_2}~~, \label{Z}
\end{equation}
and the R-R 2-form (\ref{b2}) simply reduces to
\begin{equation}
C^{(2)}=-\frac{1}{\lambda^2}\,\phi_2\,\sin {\widetilde \theta}\,
d{\widetilde \theta}\wedge d{\widetilde \varphi} ~~.
\label{b21}
\end{equation}
After these preliminaries we now study the dynamics of the $SU(N)$
gauge fields that propagate on the wrapped D5 branes. As discussed
in Refs.~\cite{d3,d3d7,ANO} for the analogue case of the
fractional branes, the low-energy action for the bosonic degrees
of freedom of the non-abelian gauge theory can be inferred from
the abelian world-volume action (\ref{borni78}) with the following
procedure: take the limit $\alpha'\to 0$ keeping fixed the
combination
\begin{equation}
\Psi = (2\pi\alpha')^{-1}\,Z
\label{Psi}
\end{equation}
that has to play the role of the complex scalar of the ${\cal
N}=2$ vector multiplet, and then promote $F$ and $\Psi$ to $SU(N)$
fields by giving them an adjoint index $A$ and replacing all
derivatives with the covariant ones~\footnote{In flat space or in
orbifold backgrounds where the classical D brane solutions can be
explicitly obtained using boundary states (see {\it e.g.}
Ref.~\cite{bs}), this procedure can be made more rigorous by
computing string scattering amplitudes as done for example in
Ref.~\cite{MERLATTI}.}. If we normalize the $SU(N)$ generators in
such a way that ${\rm tr}(T^AT^B)=(1/2)\,\delta^{AB}$ for the
fundamental representation, then the above procedure leads to
\begin{equation}
S_{YM} = -\,\frac{1}{g^{2}_{\rm YM}} \int d^4 x  \left\{ \frac{1}{4}
 F_{\alpha \beta}^A \,F^{\alpha \beta}_A + \frac{1}{2} D_{\alpha}
 {\overline\Psi}^A\, D^{\alpha} \Psi_A \right\} + \frac{\theta_{\rm
 YM}}{32 \pi^2} \int d^4 x \,F_{\alpha \beta}^A\, {\widetilde{F}}^{\alpha
 \beta}_A \label{bound53}
\end{equation}
where
\begin{eqnarray}
\frac{1}{g^2_{\rm YM}}&=&\frac{\tau_5\,(2\pi)^2\alpha'^2}{2}
\int_0^{2\pi}\!\!d{\widetilde\varphi}\int_0^\pi \!\!d{\widetilde\theta}~
{\rm e}^{-\Phi}\,H\,\sqrt{-\det G}~=~\frac{N}{4\pi^2}\,\log\rho~~,
\label{gym2} \\
\theta_{\rm YM} &=&{\tau_5\,(2\pi)^4\alpha'^2}
\int_0^{2\pi}\!\!d{\widetilde\varphi}\int_0^\pi \!\!d{\widetilde\theta}~
C^{(2)}_{\tilde\theta \tilde\varphi}~=~ -2N\,\phi_2~~.
\label{theta2}
\end{eqnarray}
In other words, the inverse square of the YM coupling constant is
proportional to the volume of the 2-sphere along which the D5
branes are wrapped computed in the ten-dimensional metric, whereas
the YM $\theta$-angle is proportional to the flux of the R-R
2-form across this 2-sphere. Notice that in deriving these results
we have used eqs.(\ref{tau5}) and (\ref{lambda}), the explicit
form of the metric (\ref{metfin}), of the dilaton (\ref{dil}) and
of the R-R 2-form (\ref{b21}), as well as the fact that
eq.(\ref{new49}) implies that $z=\log\rho$ for $\sigma=0$. In the
following two sub-sections we will use the results (\ref{gym2})
and (\ref{theta2}) to derive the $\beta$-function and the chiral
anomaly of the ${\cal N}=2$ SYM theory.

\subsection{Running Coupling Constant and $\beta$-function}

To obtain further information on the gauge theory we have to find
the precise relationship between the gravitational coordinates and
the scales of the gauge theory which, in some sense, is the
essence of the gauge/gravity correspondence. In the present case,
it is not difficult to find this relationship. In fact, let us
denote by $\mu$ the (arbitrary) mass-scale at which the theory is
defined and by $\Lambda$ the (fixed) dynamically generated scale
which is the analogue of $\Lambda_{\rm QCD}$ for quantum
chromodynamics. Under a scale transformation with parameter ${s}$
the mass $\mu$ transforms according to
\begin{equation}
\mu~\to~s\,\mu
\label{mus}
\end{equation}
but the physical mass $\Lambda$ remains fixed. This is the signal of the
lack of conformal invariance in this theory. On the other hand, it is
known that the complex scalar $\Psi$ of the ${\cal N}=2$ vector multiplet
has a (protected) mass-dimension 1, that is
\begin{equation}
\Psi~\to~s\,\Psi~~.
\label{psis}
\end{equation}
If we now use eqs.(\ref{Psi}) and (\ref{Z}), we easily conclude
that the scale transformation (\ref{psis}) implies
\begin{equation}
\rho~\to~s\,\rho
\label{rhos}
\end{equation}
{\it i.e.} in the dual gravitational description the scale
transformation of the gauge theory is realized as a dilatation of
the 2-plane in the transverse space of the wrapped D5 branes.
Combining the fact that $\rho$ is dimensionless with the
transformations (\ref{mus}) and (\ref{rhos}), we are lead the
following identification
\begin{equation}
\rho = \frac{\mu}{\Lambda}~~.
\label{holo2}
\end{equation}
This is a sort of ``holographic'' relation, at least in an extended sense,
because it establishes a correspondence
between a quantity of the gravity theory (the coordinate $\rho$) and
a quantity of the gauge theory (the scale $\mu$) even if
it does not follow from a standard holographic bulk/boundary relation.
The fact that $\rho$ and $\mu$ are directly proportional to each other
can also be established by looking at the energy of a string that stretches
up to a distance $\rho$ in the transverse plane. The energy $E$ of
such a string is proportional to its world-volume
per unit time \cite{Peet,HKO}, namely
\begin{equation}
E\sim \int^\rho \sqrt{-G_{00}\,G_{\rho\rho}}~ d\rho \sim \rho
\label{energy}
\end{equation}
where we have used the metric (\ref{metfin}).
This energy is the natural scale to use in order
to regulate the theory and hence it is natural to take $E\sim \mu$,
which in turn leads to (\ref{holo2}).
Notice that this relation implies that the UV (IR) regime of the
gauge theory corresponds to the large (small) distance region in the
dual gravitational description.

If we insert (\ref{holo2}) into (\ref{gym2}) we can obtain
the expression of the running coupling constant at the scale $\mu$,
namely
\begin{equation}
\frac{1}{g^2_{\rm YM}(\mu)} = \frac{N}{4\pi^2}\,\log\frac{\mu}{\Lambda}~~,
\label{running2}
\end{equation}
from which we can derive the $\beta$-function
\begin{equation}
\beta(g_{\rm YM}) = -\frac{N}{8\pi^2}\,g_{\rm YM}^3~~.
\label{beta2}
\end{equation}
This is the correct field-theory result, numerical factors
included. We would like to stress that this $\beta$-function has
been obtained without using the probe approach which instead is
appropriate to describe the theory in the Coulomb branch (for a
recent review, see Ref.~\cite{REV})~\footnote{The $\beta$-function
for the ${\cal N}=2$ SYM theory in the Coulomb branch has been
first derived from the wrapped D5 brane solution in
Ref.~\cite{dario} up to some numerical factors.}.

\subsection{Chiral Anomaly}

We now discuss the chiral anomaly from the dual gravitational
point of view. It is well known that the ${\cal N}=2$ SYM theory
possesses at the classical level an abelian $U(1)_R$ symmetry
which becomes anomalous at the quantum level. For definiteness, we
fix the $R$-charge of the complex scalar $\Psi$ to be 2, so that
the gauginos have the conventional $R$-charge 1. This means that
under a chiral transformation with parameter $\varepsilon$ we have
\begin{equation}
\Psi ~\to~{\rm e}^{2{\rm i}\varepsilon}\,\Psi \label{rpsi}~~,
\end{equation}
and that the chiral anomaly corresponds to the following
transformation of the $\theta$-angle
\begin{equation}
\theta_{\rm YM} ~\to~\theta_{YM} - 4N\,\varepsilon~~.
\label{anomaly2}
\end{equation}
{F}rom eqs.(\ref{Psi}) and (\ref{Z}) it is natural to interpret
the transformation (\ref{rpsi}) as due to a shift in the angle
$\phi_2$, namely
\begin{equation}
\phi_2~\to~\phi_2 + 2\,\varepsilon \label{shiftpsi}~~,
\end{equation}
which is also consistent with the anomaly transformation
(\ref{anomaly2}) in view of eq.(\ref{theta2}).

Thus, the chiral $U(1)_R$ transformations of the gauge theory are
realized in the dual gravitational description as phase rotations
in the transverse plane where the branes can move with a
multiplicity factor (2 in our case) that depends on the charge
assignments. As one can easily see from the explicit expressions
(\ref{dil}), (\ref{b2}) and (\ref{metfin}), the transformation
(\ref{shiftpsi}) is an isometry of the metric but it is not a
symmetry of the full solution since the R-R 2-form $C^{(2)}$
explicitly depends on the angle $\phi_2$ and is not invariant
under (\ref{shiftpsi}). This fact is the gravitational counterpart
of the occurrence of a chiral anomaly in the dual gauge
theory~\cite{KOW,ANO}. However, not all different values of the
R-R 2-form $C^{(2)}$ are physically different: in fact what really
matters is the value of the flux across the 2-sphere
\begin{equation}
\frac{1}{4\pi^2\alpha'g_s}\int_{S_2} \!C^{(2)}
=-\,\frac{N\,\phi_2}{\pi}\label{flux}
\end{equation}
which is allowed to change by integer values. Thus, the following
shifts
\begin{equation}
\phi_2~\to~\phi_2 + \frac{\pi}{N}\,k \label{shiftk}
\end{equation}
(with $k$ integer) are true symmetries of the supergravity
solution. On the gauge theory side, these shifts correspond to the
non-anomalous ${\mathbf Z}_{4N}$ rotations with parameter
$\varepsilon=(\pi/2N)\,k$, which change the $\theta$-angle by
integer multiples of $2\pi$, as one can easily see from
(\ref{anomaly2}).

\vskip 1.5cm
\section{The ${\cal N}=1$ Super Yang-Mills Theory}

The supergravity solution presented in subsection 2.2 describes
the geometry produced by $N$ D5 branes wrapped on a supersymmetric
2-cycle in such a way that one eighth of the 32 supercharges of
Type IIB are preserved. The four-dimensional unwrapped part of the
D5 world-volume is the Minkowski space-time where a supersymmetric
gauge theory with 4 supercharges is defined. By using the methods
of Ref.~\cite{sodov}, one can see that the massless open string
modes that propagate in this Minkowski space fill a ${\cal N}=1$
vector multiplet in the adjoint representation of $SU(N)$, and
thus the low-energy four-dimensional gauge theory is a pure ${\cal
N}=1$ SYM theory with gauge group $SU(N)$.

We now show how to use the supergravity solution
(\ref{stri987})-(\ref{f3exact}) to extract information on the corresponding
gauge theory. To this aim, we consider again the world-volume action
(\ref{borni78}) of a wrapped five-brane with a gauge field
strength $F$ and then extract from it the quadratic terms in $F$.
In contrast to the ${\cal N}=2$ case discussed in the previous
section, this time there is no direction in which the branes can
move, since the entire transverse space is curved as a consequence
of the topological twist of the normal bundle that has been
performed. Thus, no scalars survive and the non-abelian bosonic
action that is obtained with the procedure discussed in Section 3,
is simply
\begin{equation}
S_{YM} = -\,\frac{1}{4g^{2}_{\rm YM}} \int d^4 x\,
 F_{\alpha \beta}^A \,F^{\alpha \beta}_A  + \frac{\theta_{\rm
 YM}}{32 \pi^2} \int d^4 x \,F_{\alpha \beta}^A\, {\widetilde{F}}^{\alpha
 \beta}_A \label{bound99}
\end{equation}
where
\begin{eqnarray}
\frac{1}{g^2_{\rm YM}}&=&\frac{\tau_5\,(2\pi)^2\alpha'^2}{2}
\int_0^{2\pi}\!\!d{\widetilde\varphi}\int_0^\pi
\!\!d{\widetilde\theta}~ {\rm e}^{-3\Phi}\,\sqrt{-\det G}~~,
\label{gym1} \\ \theta_{\rm YM} &=&{\tau_5\,(2\pi)^4\alpha'^2}
\int_0^{2\pi}\!\!d{\widetilde\varphi}\int_0^\pi
\!\!d{\widetilde\theta}~ C^{(2)}_{\tilde\theta \tilde\varphi}~~.
\label{theta1}
\end{eqnarray}
Exactly as in the ${\cal N}=2$ case (see eqs.(\ref{gym2}) and (\ref{theta2}))
the inverse square of the YM coupling constant is proportional to the volume
of the 2-sphere on which the D5 branes are wrapped and the YM vacuum
angle is proportional
to the flux of the R-R 2-form across this 2-sphere. However, we
would like to stress that our definition (\ref{gym1}) of the YM
coupling constant is different from the one used in Ref.~\cite{MN}
and several other papers that followed afterwards, because we compute the
volume of the 2-sphere using the ten-dimensional metric of the wrapped
D5 branes, and not the metric (\ref{anme}) of the domain-wall solution
of the seven-dimensional gauged supergravity. Our definition is the
natural one from a string-theory point of view; on the other hand, no one
doubts that the YM $\theta$-angle is related to the flux of the
R-R form in ten-dimensions, and thus it appears more
acceptable that also the other parameter of the SYM action be related
to quantities of the ten-dimensional solution as in (\ref{gym1}).
The difference between our definition of the YM constant and the one of
Ref.~\cite{MN} will have interesting consequences for the gauge
theory, especially in the IR regime.

If we insert the explicit form of the supergravity
solution (\ref{stri987})-(\ref{f3exact})
in eq.(\ref{gym1}), after simple calculations
we obtain
\begin{equation}
\frac{1}{g_{YM}^{2}} = \frac{N}{32\pi^2}\,Y(\rho)\int_{0}^{\pi}
\! d {\widetilde{\theta}}~  \sin
{\widetilde{\theta}}
\left[\,1 + \frac{\cot^2\!
{\widetilde{\theta}} }{Y(\rho)}\,\right]^{1/2}
\label{gym84}
\end{equation}
with
\begin{equation}
Y(\rho) = 4\,{\rm e}^{2h(\rho)}+{a(\rho)^2}=4\rho\,\coth 2\rho -1
\label{Y}
\end{equation}
where in the last step we have used eqs.(\ref{solh6}) and
(\ref{solh62}).
It is interesting to observe that the right hand side of
eq.(\ref{gym84}) can be written in terms of the complete elliptic integral
of the second kind
\begin{equation}
E(x) \equiv \int_{0}^{\pi/2}\!\! d\phi~ \sqrt{1-x^2\,\sin^2\!\phi}~~;
\label{elliptic}
\end{equation}
in fact, we find
\begin{equation}
\frac{1}{g_{YM}^2} = \frac{N\,Y(\rho)}{16\pi^2}\,
E\left(\sqrt{\frac{Y(\rho)-1}{Y(\rho)}}\right)~~.
\label{gymfin}
\end{equation}
Using the properties of the elliptic integral, it is easy to see that
\begin{eqnarray}
\frac{1}{g_{\rm YM}^2}&\simeq&\frac{N\,\rho}{4\pi^2}~~~~~~~~~{\rm for}
~~\rho\to\infty~~,\label{gymUV}\\
\frac{1}{g_{\rm YM}^2}&\simeq&\frac{N}{32\pi}~~~~~~~~~{\rm for}~~\rho\to 0
~~.\label{gymIR}
\end{eqnarray}
The large-$\rho$ behavior (\ref{gymUV}) is the same as in Ref.~\cite{MN},
but the small-$\rho$ behavior (\ref{gymIR}) is very different, since we get
a {\it finite} coupling at $\rho=0$, in contrast to the divergent
one of Ref.~\cite{MN}.

Finally, inserting in eq.(\ref{theta1}) the classical profile of
the R-R 2-form (\ref{f3exact}), we find
\begin{equation}
\theta_{\rm YM} = -\frac{N}{2} \,\psi ~~.\label{tg1}
\end{equation}
In the following we will exploit these results to discuss some
relevant features of the pure ${\cal N}=1$ SYM theory from the gravitational
point of view.

\subsection{Chiral Anomaly}
As is well-known, the ${\cal N}=1$ SYM theory possesses a
classical abelian $U(1)_R$ symmetry which becomes anomalous at the
quantum level. If we assign $R$-charge 1 to the gaugino
$\lambda(x)$ so that under a chiral transformation with parameter
$\varepsilon$
\begin{equation}
\lambda(x)~\to~{\rm e}^{{\rm i}\,\varepsilon}\,\lambda(x)~~,
\label{gaugin}
\end{equation}
then the presence of the anomaly implies that
\begin{equation}
\theta_{\rm YM} ~\to~\theta_{\rm YM} - 2N\varepsilon~~.
\label{anom1}
\end{equation}

{F}rom eq.(\ref{tg1}) it is clear that the chiral transformations
of the SYM theory must be realized in the gravitational
description as shifts in the angle $\psi$. The only question is
which is the multiplicity factor. Now we argue that the correct
transformation law of $\psi$ is
\begin{equation}
\psi~\to~\psi+4\,\varepsilon~~, \label{psiepsi}
\end{equation}
which indeed allows to obtain the anomaly rule (\ref{anom1})
directly from (\ref{tg1}). The factor of 4 in (\ref{psiepsi}) has
a natural explanation: in fact it is $\psi/2$ that is the
appropriate angular variable with period $2\pi$ which must
transform with multiplicity 2, just like the angle $\phi_2$ in the
${\cal N}=2$ case (see eq.(\ref{shiftpsi})). This argument can be
made more rigorous by observing that there is a simple relation
between the Euler angles that we used to parameterize the 3-sphere
in the up-lifting process of the ${\cal N}=1$ solution and the
angles that instead we used in the up-lift of the ${\cal N}=2$
solution. In particular, as is discussed in the Appendix,
$\psi=\phi_1+\phi_2$ (see (\ref{para99})), and since  under a
chiral transformation of parameter $\varepsilon$ both $\phi_1$ and
$\phi_2$ shift by $2\,\varepsilon$, then eq.(\ref{psiepsi})
immediately follows.

The transformation (\ref{psiepsi}) is {\it not} a symmetry of the
supergravity solution (\ref{stri987})-(\ref{f3exact}) and is not
even an isometry of the metric. This situation is thus very
different from the ${\cal N}=2$ case discussed in the previous
section. However, there is a region where the transformation
(\ref{psiepsi}) is an isometry, namely the large-$\rho$ region
where the function $a(\rho)$ becomes exponentially small and can
be neglected. In fact, if we remove $a$, then all explicit $\psi$
dependence disappears from the metric (\ref{stri987}) but still
remains in the R-R 2-form (\ref{f3exact}), which therefore is not
invariant under (\ref{psiepsi}). However, as we discussed in
Section 3.2, the relevant quantity that should be considered is
the flux of $C^{(2)}$ across the 2-sphere, {\it i.e.}
\begin{equation}
\frac{1}{4\pi^2\alpha'g_s}\int_{S_2} \!\left.C^{(2)}\right|_{a=0}
=-\,\frac{N\,\psi}{4\pi}\label{flux1}
\end{equation}
which is allowed to change by integer values. Thus the
transformations (\ref{psiepsi}) with $\varepsilon=(\pi/N)k$ and $k$
integer are true symmetries of the supergravity background in the 
large-$\rho$ region. These are precisely 
the non-anomalous ${\mathbf Z}_{2N}$
transformations of the gauge theory that correspond to shifts of the
$\theta$-angle by integer multiples of $2\pi$. What we have described
here is therefore the gravitational counterpart of the well-known breaking
of $U(1)_R$ down to ${\mathbf Z}_{2N}$.

However, in the true supergravity solution $a(\rho)$ is not
vanishing and thus even the non-anomalous ${\mathbf Z}_{2N}$
transformations are not symmetries of the background. In fact,
both in the metric and in the R-R 2-form, $a(\rho)$ appears in
front of terms that explicitly involve $\cos \psi$ and $\sin\psi$,
and these are clearly invariant only under shifts of $2\pi$. Thus,
the only non-anomalous symmetries of the supergravity background
are given by (\ref{psiepsi}) with $\varepsilon=k\pi$ and $k$
integer. This phenomenon is the gravitational counterpart of the
spontaneous breaking of the chiral symmetry from ${\mathbf Z}_{2N}
~\to~{\mathbf Z}_2$.

\subsection{Gaugino Condensate}

In the previous subsection we have seen that the presence of a
non-vanishing $a(\rho)$ in the supergravity solution is
responsible for the spontaneous chiral symmetry breaking to
${\mathbf Z}_2$, which, on the gauge theory side, is accompanied
by the presence of a non-vanishing value of the gaugino condensate
$\langle \lambda^2 \rangle\equiv\langle 0|\left(\frac{{\rm
Tr}\,\lambda^2}{16\pi^2}\right)|0\rangle$. Thus, it appears very
natural to conjecture that the gravitational dual of this
condensate is precisely the function $a(\rho)$ that is present in
the supergravity solution (\ref{stri987})-(\ref{f3exact}). As a
matter of fact, this idea has been made more precise in
Ref.~\cite{MILANO} where a direct relation between $\langle
\lambda^2 \rangle$ and $a(\rho)$ has been established by adapting
the techniques of the AdS/CFT correspondence to the wrapped
branes.

The gaugino condensate $\langle \lambda^2\rangle$ belongs to a
class of gauge invariant operators which do not acquire any
anomalous dimensions. This happens because the gaugino condensate
is the lowest component of the so-called anomaly multiplet whose
scale dimensions are protected by virtue of the fact that its top
component is the trace of the energy-momentum tensor which is a
conserved current. Thus, since the engineering dimension of
$\langle \lambda^2 \rangle$ is 3, we have
\begin{equation}
\langle \lambda^2 \rangle = c\,\Lambda^3 \label{condens}
\end{equation}
where $\Lambda$ is the (exact) dynamical scale of the ${\cal N}=1$
SYM theory and $c$ a computable constant. Several explicit
calculations lead to $c=1$ \cite{gaugino}.

In view of the findings of Ref.~\cite{MILANO} and of our previous
discussion, we now propose to identify the function $a(\rho)$ given in
(\ref{solh62}) with the gaugino condensate $\langle \lambda^2
\rangle$ measured in units of the (arbitrary) mass scale $\mu$
that is introduced to regulate the theory. Thus, we write
\begin{equation}
a(\rho) = \frac{\Lambda^3}{\mu^3}~~. \label{holo1}
\end{equation}
This crucial equation allows us to establish a precise relation
between the supergravity radial coordinate $\rho$ and the scales
of the gauge theory, and is the strict analogue of the
``holographic'' relation (\ref{holo2}) of the ${\cal{N}}=2$ model.
Notice that since $a(\rho)\to 0$ for $\rho\to\infty$, once again
the large-$\rho$ region corresponds to the UV regime of the gauge
theory, and vice-versa the small-$\rho$ region corresponds to the
IR regime. 
In the following we will exploit the relation (\ref{holo1}) to
compute the perturbative $\beta$-function of the pure
${\cal{N}}=1$ SYM theory and get some insights on non-perturbative
corrections.

\subsection{Running Coupling Constant and $\beta$-function}

The gauge coupling constant has been defined in (\ref{gym84}) as a
function of the radial coordinate $\rho$ which, in turn, is
related to the scales of the gauge theory as prescribed by
(\ref{holo1}). Thus, by combining these two equations and
eliminating the $\rho$ dependence, we could in principle obtain an
exact expression for the running coupling constant of the
${\cal{N}}=1$ SYM theory. Unfortunately, it appears very difficult
to manipulate eqs. (\ref{gym84}) and (\ref{holo1}) in an analytic
way and exhibit the running coupling constant in a closed form.
However, despite this fact, many interesting results can still be
derived.

First of all, if we use in (\ref{holo1}) the explicit expression
of $a(\rho)$ given in (\ref{solh62}), we easily obtain
\begin{equation}
\frac{\partial \rho}{\partial \log (\mu/\Lambda)} =
\frac{3}{2}\left[\frac{1}{1-(2\rho)^{-1}+2\,{\rm e
}^{-4\rho}\,\left(1-{\rm
e}^{-4\rho}\right)^{-1}}\right]~~.\label{rholog}
\end{equation}
The right hand side has a nice interpretation in the large-$\rho$
region, {\it i.e.} in the UV. In fact, for $\rho\to\infty$ the
fraction in square brackets receives two types of contributions:
one from terms that are negative powers of $\rho$, and the other
from terms that involve also powers of ${\rm e}^{-4\rho}$. If we
recall that in the UV region $\rho$ is directly proportional to
the inverse square of the gauge coupling constant as shown in
(\ref{gymUV}), it is very easy to realize that the first type of
terms corresponds to perturbative loop contributions while the
second describes non-perturbative instanton-like effects.
Actually, we can be more precise and write explicit formulas. Let
us first concentrate on the power-like part and neglect all terms
that vanish exponentially for $\rho\to\infty$. Then, if we use
the leading UV asymptotic behavior
(\ref{gymUV}) and thus trade $\rho$ for $4\pi^2/(N\,g_{\rm YM}^2)$, on
the one hand we have
\begin{equation}
\frac{\partial \rho}{\partial \log (\mu/\Lambda)} =
-\frac{8\pi^2}{Ng_{\rm YM}^3}\,\beta(g_{\rm YM}) \label{beta}
\end{equation}
where $\beta(g_{\rm YM})$ is the $\beta$-function, and on the
other hand from (\ref{rholog}) we have
\begin{equation}
\frac{\partial \rho}{\partial \log (\mu/\Lambda)} = \frac{3}{2}
\left[{1-\frac{Ng_{\rm YM}^2}{8\pi^2}}\right]^{-1}~~.
\label{rholog1}
\end{equation}
Combining these two equations, we obtain
\begin{equation}
\beta(g_{\rm YM})=-\frac{3Ng_{\rm YM}^3}{16\pi^2}
\left[{1-\frac{Ng_{\rm YM}^2}{8\pi^2}}\right]^{-1}~~.
\label{betaUV}
\end{equation}
This is the complete perturbative NSVZ $\beta$-function
of the pure ${\cal{N}}=1$ SYM theory with gauge group $SU(N)$ in
the Pauli-Villars regularization scheme~\cite{NSVZ}. It is remarkable to
see that a classical supergravity solution representing wrapped D5
branes is able to completely reproduce it!

It is interesting to observe that the 1-loop approximation
corresponds to keep in the supergravity solution only the leading
terms in the $\rho\to\infty$ expansion. In fact, if in
(\ref{rholog}) we approximate the square bracket to 1, we get
$\rho\sim\frac{3}{2}\log({\mu}/{\Lambda})$ which is indeed the
correct relation between the radial coordinate and the scale
$\Lambda$ at 1-loop~\cite{MILANO}. If instead we keep also the
$\rho^{-1}$ term in (\ref{rholog}), we obtain the complete
perturbative result (\ref{rholog1}) and the NSVZ $\beta$-function.
However, the full supergravity solution contains more than this,
since there are also terms that vanish exponentially for large
$\rho$ and correspond to non-perturbative effects. Keeping these
terms in (\ref{rholog}) amounts to replace the expression inside
the square brackets of (\ref{rholog1}) and (\ref{betaUV}) with
\begin{equation}
1-\frac{Ng_{\rm YM}^2}{8\pi^2}+\frac{2\,\exp\Big({-\frac{16\pi^2}{N
g_{\rm YM}^2}}\Big)}{1-
\exp\Big({-\frac{16\pi^2}{N
g_{\rm YM}^2}}\Big)}~~. \label{den}
\end{equation}
From this formula we see that the non-perturbative corrections
have the form of instantons with fractional charge $2k/N$ where
$k$ is a positive integer. The fractional instantons have recently
been shown in Ref.~\cite{khoze} to play a crucial role in the pure
${\cal N}=1$ SYM theory since they are the elementary field
configurations that directly contribute to the gaugino condensate.
It would be very interesting to check with a field theory analysis
whether they also contribute to the $\beta$-function as our
supergravity analysis suggests.

We conclude by observing that the presence of these non-perturbative
effects modifies the running of the coupling constant which remains finite
even at low energy. In fact, from (\ref{holo1}) we see that the IR limit
$\mu\to \Lambda$ corresponds to the limit $\rho\to 0$ in which the
coupling constant smoothly tends to a finite value (see eq. (\ref{gymIR})).
This smooth behaviour and the absence of a Landau pole 
qualitatively resembles
the soft confinement scenario of QCD.

\subsection{Instantons}

It well-known that a system of $N$ D3 branes and $k$ D(-1) branes
in flat space describes the $k$-instanton sector of the maximally
supersymmetric $SU(N)$ Yang-Mills theory in four dimensions
\cite{douglas}. This idea has been successfully generalized to
other less supersymmetric configurations by considering, for
example, systems of D3/D(-1) fractional branes in orbifold
backgrounds. Thus, it appears very natural to think that the same
happens also in the context of wrapped branes. The idea is then to
consider a system of $N$ D5 branes and $k$ D1 branes wrapped on
the same supersymmetric 2-cycle and see whether the $k$ wrapped
D-strings branes account for the $k$ instanton contributions. This
analysis has been performed in Ref.~\cite{LS} where, however,
negative conclusions have been reached. We now show that this is
not the case and that, also in the wrapped brane case, the
instantons are correctly represented by D branes with four
dimensions less than the branes which support the gauge degrees of
freedom.

The world-volume action of a Euclidean D1 brane in the string frame
is given by
\begin{equation}
S = \tau_1 \int d^2 \!\xi ~{\rm e}^{- \Phi} \sqrt{\det  G} ~-\,
{\rm i}\,\tau_1\! \int \! C^{(2)}\label{bornd1}
\end{equation}
where
\begin{equation}
\tau_1=\frac{1}{2\pi g_s\alpha'}
\label{tau1}
\end{equation}
and the bulk fields are understood to be the pull-backs onto the
brane world-volume which is parameterized
$\xi=\{\widetilde\theta,\widetilde\varphi\}$. If the D1 brane is
in the background geometry produced by $N$ wrapped D5 branes, we
must use in the above action the metric, the dilaton and the R-R
2-form of the solutions reviewed in Section 2. In particular in
the ${\cal N}=1$ case, by using the explicit form of the metric
given in (\ref{stri987}) we get
\begin{equation}
\sqrt{\det G} = \frac{{\rm e}^\Phi}{4\lambda^2}\,
Y(\rho)\,\sin\widetilde\theta\,\sqrt{1+\frac{\cot^2\widetilde\theta}
{Y(\rho)}}
\label{metd1}
\end{equation}
where $Y(\rho)$ is defined in (\ref{Y}). Inserting this result
into the action (\ref{bornd1}), recalling from (\ref{f3exact})
that the relevant component of the R-R 2-form is
\begin{equation}
C^{(2)}_{\tilde \theta \tilde \varphi} = -\,\frac{1}{4
\lambda^2}\, \psi\, \sin {\widetilde{\theta}}~~, \label{rr2d1}
\end{equation}
and performing the integrals over the world-volume coordinates, we
find
\begin{equation}
S= \tau_1\,\frac{\pi}{\lambda^2}\,Y(\rho)
\,E\left(\sqrt{\frac{Y(\rho)-1}{Y(\rho)}}\right) +\,{\rm
i}\,\tau_1\,\frac{\pi\,\psi}{\lambda^2}~~. \label{instaction}
\end{equation}
Finally, if we express $\tau_1$ and $\lambda$ in terms of the
string parameters as in (\ref{tau1}) and (\ref{lambda1}), and use
the definitions of the YM coupling constant and the $\theta$-angle
given in (\ref{gymfin}) and (\ref{tg1}), it is easy to see that
\begin{equation}
S= \frac{8\pi^2}{g_{\rm YM}^2}-\,{\rm i}\,\theta_{\rm YM}
\label{instaction1}
\end{equation}
which is the correct form of the instanton action! Clearly, to
obtain the $k$-instanton action we should consider $k$ wrapped D1
branes in the background of $N$ wrapped D5 branes, which amounts
simply to multiply the result (\ref{instaction1}) by $k$.

Therefore, we conclude that the
gauge theory instantons of the ${\cal N}=1$ SYM theory are indeed
represented by D1 branes that wrap the same supersymmetric 2-cycle
of the D5 branes, as one should expect from general
considerations~\footnote{The discrepancies with Ref.~\cite{LS}
originate from a different choice of normalization for the $SU(N)$
generators as well as from a different definition of the YM
coupling constant that for us is proportional to the volume of the
2-cycle computed with the ten-dimensional metric and not with the
seven-dimensional domain-wall metric as in Ref.~\cite{LS}.}. We
also notice that the above calculations can be performed in the
${\cal N}=2$ SYM theory using the supergravity solution of Section
2.1 which leads to similar results.

\section{Conclusions}

We have shown that many interesting features of the pure ${\cal
N}=1$ and ${\cal N}=2$ SYM theories in four dimensions are
quantitatively encoded in the supergravity solutions that describe
D5 branes wrapped on supersymmetric 2-cycles. These features
comprise the running of the gauge coupling constant, the
$\beta$-function, the chiral anomaly and instantons. In the ${\cal
N}=1$ case we have also discussed the gravitational dual of the
gaugino condensate and exploited it to obtain the complete NSVZ
$\beta$-function. Moreover we have found the occurrence of
non-perturbative corrections in the form of fractional instantons
with topological charge $2/N$ that smooth out the running of the
coupling constant which never diverges. The precise meaning of the
fractional instanton corrections and their relevance for the IR
physics deserve, however, further study from a field theory point
of view. Finally, we observe that in the ${\cal N}=2$ case the
supergravity analysis that we have presented allows to obtain only
the perturbative part of the $\beta$-function and no instanton
corrections. In this respect it is worth pointing out that,
differently from the ${\cal N}=1$ case, the ${\cal N}=2$
supergravity solution exhibits a naked singularity at small
distances where it becomes unphysical. It would very interesting
to see whether it is possible to resolve this singularity and
eventually find the known non-perturbative instanton corrections
of the ${\cal N}=2$ SYM theory from a resolved supergravity
solution. Another interesting development would be to explore if
the MN solution and the results we have found here can be used to
quantitatively investigate other features of the ${\cal N}=1$ SYM
theory and if they can be extended also to theories in which
supersymmetry is completely broken.

\vskip 1.5cm \noindent {\large {\bf Acknowledgments}} \vskip 0.2cm
\noindent We thank R. Auzzi, M. Bertolini, M. Bill\`o, M. Frau, K. Konishi,
E. Imeroni, L. Magnea, R. Marotta and I. Pesando for useful discussions and
exchange of ideas.
\vskip 1.5cm
\appendix
\section{Parameterizations of the 3-sphere}

There are several ways to describe a 3-sphere $S^3$. Here we
briefly describe the two parameterizations we have used in Section
2 to up-lift the domain wall-solution from seven to ten
dimensions.

The defining equation for a unit 3-sphere is
\begin{equation}
\mu_1^2+\mu_2^2+\mu_3^2+\mu_4^2=1 \label{sphere}
\end{equation}
One explicit parameterization of the $\mu_i$'s is the following
\[\mu_1 = \cos \theta \cos \phi_1  ~~,~~\mu_2 = \cos \theta \sin \phi_1~~,
\]
\begin{equation}
\mu_3 = \sin \theta \cos \phi_2 ~~,~~\mu_4 = \sin \theta \sin
\phi_2~~, \label{sphere3}
\end{equation}
with $ 0\leq \theta  \leq \frac{\pi}{2} $ and $ 0 \leq\phi_{1,2}
\leq 2 \pi$. In this parameterization, the metric of the 3-sphere
is
\begin{equation}
ds^2=\sum_{i=1}^4
d\mu_i^2=d\theta^2+\cos^2\theta\,d\phi_1^2+\sin^2\theta
\,d\phi_2^2~~.
\end{equation}
These are the coordinates that we used in Section 2.1 in order to
up-lift to ten dimensions the domain-wall solution in the ${\cal
N}=2$ case.

A second convenient parameterization of the $\mu_i$'s is in terms
of the Euler angles, namely
\[ \mu_3 = \cos \frac{\theta'}{2}\,
\cos \frac{\psi + \phi}{2}~~,~~ \mu_4 = \cos \frac{\theta'}{2}
\,\sin \frac{\psi + \phi}{2}~~,
\]
\begin{equation}
\mu_2 = \sin \frac{\theta'}{2}\, \cos \frac{\psi - \phi}{2}~~,~~
\mu_1 = \sin \frac{\theta'}{2}\, \sin \frac{\psi - \phi}{2}~~,
\label{para49}
\end{equation}
with $0\leq \theta'\leq\pi$, $0\leq\phi\leq 2\pi$ and
$0\leq\psi\leq 4\pi$. The corresponding left-invariant 1-forms of
$S^3$ are
\[
\sigma^1 =  \frac{1}{2}\Big[\cos \psi \,d\theta' + \sin \theta'
\sin \psi \,d \phi  \Big]~~,~~\sigma^2 = -\frac{1}{2} \Big[\sin
\psi \,d \theta' - \sin \theta' \cos \psi \,d\phi \Big]~~,
\]
\begin{equation}
\sigma^3 = \frac{1}{2} \Big[d \psi + \cos \theta' \,d \phi
\Big]~~, \label{diffe49}
\end{equation}
which close the $SU(2)$ differential algebra
\begin{equation}
d\sigma^a = -\epsilon^{abc} \sigma^b\wedge \sigma^c~~.
\end{equation}
In this parameterization the metric of the unit 3-sphere is
\begin{equation}
ds^2=\sum_{i=1}^4 d\mu_i^2=\sum_{a=1}^3
(\sigma^a)^2=\frac{1}{4}\Big[ (d\psi + \cos\theta'
\,d\phi)^2+\sin^2\theta' \,d\phi^2+d\theta'^2\Big]~~.
\end{equation}
These are the coordinates that we used in Section 2.2 in order to
up-lift to ten dimensions the domain-wall solution in the ${\cal
N}=1$ case.

By comparing (\ref{sphere3}) and (\ref{para49}), it is immediate
to see that the two sets of angles are related as follows
\begin{equation}
\theta'=2\theta~~~,~~~\phi=\phi_1-\phi_2~~~,~~~\psi=\phi_1+\phi_2~~.
\label{para99}
\end{equation}


\end{document}